\documentclass	[prb,twocolumn,showpacs]{revtex4-1}

\usepackage{graphicx}
\usepackage{dcolumn}
\usepackage{bm}
\usepackage{amsmath}
\usepackage{epstopdf}
\usepackage{color}
\usepackage{lipsum}
\begin{document}

\title{Spin-Wave Spectrum in Magnetic Nanodot with Continuous Transition between Vortex, Bloch-type Skyrmion and N\'eel-type Skyrmion States}
\author{M.~Mruczkiewicz$^{1}$ } \email{m.mru@amu.edu.pl}
\author{M.~Krawczyk$^{2}$ } \email{krawczyk@amu.edu.pl}
\author{K. Y. Guslienko$^{3,4}$ } \email{}
\affiliation{$^{1}$Institute of Electrical Engineering, Slovak Academy of Sciences, Dubravska cesta 9, 841 04 Bratislava, Slovakia\\
$^{2}$Faculty of Physics, Adam Mickiewicz University in Poznan, Umultowska 85, Pozna\'{n}, 61-614, Poland \\
$^{3}$Depto. Fisica de Materiales, Universidad del País Vasco, UPV/EHU, 20018 San Sebastian, Spain \\
$^{4}$IKERBASQUE, the Basque Foundation for Science, 48013 Bilbao, Spain}

\date{\today}

\begin{abstract}

We study spin-wave excitations in a circular ferromagnetic nanodot in different inhomogeneous, topologically non-trivial magnetization states, specifically, vortex and skyrmion states. Gradual change in the strength of the out-of-plane magnetic anisotropy and the Dzyaloshinskii-Moriya exchange interaction leads to continuous phase transitions between different stable magnetic configurations and allows for mapping of dynamic spin modes in and between the vortex, Bloch-type skyrmion and N\'eel-type skyrmion states. Our study elucidates the connections between gyrotropic modes, azimuthal spin waves and breathing modes in various stable magnetization states and helps to understand the rich spin excitation spectrum on the skyrmion background.

\end{abstract}

\maketitle

\section{Introduction}

Theory of 2D magnetic topological solitons was developed in 1980-years, see Ref. [\onlinecite{kosevich1990magnetic}] and references there. The first kind of the topological solitons, magnetic vortex stabilized in flat soft magnetic particles (dots) was discovered in 2000 by Shinjo et al. [\onlinecite{shinjo2000magnetic}]. The spin excitation spectrum over the vortex ground state was actively studied during the last decade and is now well established.\cite{guslienko2002eigenfrequencies, guslienko2006magnetic, 112,mamica2013effects} The classification of the excitations on high frequency spin waves (SWs) and low frequency vortex gyrotropic modes based on the mode symmetry and number of nodes in the dynamical magnetization profiles along the radial and azimuthal directions has been suggested. The vortex spin excitations are interesting from physical point of view but also have important potential applications, for instance, for understanding of the switching of the vortex core polarization,\cite{guslienko2001magnetization, guslienko2008dynamic} an effect that  might find an application in novel magnetic logic or memory devices.\cite{114}

2D hexagonal lattices of magnetic skyrmions (another kind of the topological solitons) were found in 2009 in thin films of some compounds with B20 cubical crystal structure without an inversion center (like MnSi, FeGe, etc.) and in some multiferroics (Cu$_{2}$SeO$_{3}$, etc.) at low temperatures. These skyrmion lattices are stabilized due to the antisymmetric Dzyaloshinskii-Moriya exchange interaction (DMI). The spin excitation spectra of the skyrmion lattices were simulated \cite{38} and measured  by broadband ferromagnetic resonance.\cite{57, schwarze2015universal} Very recently the single skyrmions stabilized in ultrathin magnetic films and dots by an interface induced DMI were discovered using X-ray imaging.\cite{moreau2016additive, boulle2016room, woo2016observation} Such individual skyrmions attracted attention due to their existence at room temperature, small (zero) magnetic field stability and high mobility in response to spin polarized current.\cite{sampaio2013nucleation} The study of spin dynamics in the skyrmion state in restricted geometry (magnetic dots, stripes etc.) is equally important, however, it is still in the beginning stage. The theory of the topological soliton dynamics developed so far, concerns so called precession solitons in infinite 2D systems (thin magnetic films). Typically the Belavin-Polyakov (no magnetic anisotropy) solitons\cite{ivanov1999soliton} and the precessional solitons in uniaxial "easy" axis ferromagnets \cite{sheka2001internal} were considered. The need to consider precession of the magnetic solitons appeared due to the problem with the soliton stability in the absence of the high-order exchange or Dzyaloshinskii-Moriya exchange interaction. The solitons were made conditionally stable fixing the number of bounded magnons.\cite{kosevich1990magnetic} The situation with magnetic dots/stripes is different: the topological soliton can be stabilized either by the DMI (bulk or interface) or magnetostatic interactions at some finite value of the uniaxial magnetic anisotropy. There are zero-frequency spin excitation modes for the topological solitons in infinite films (radially symmetric breathing mode and azimuthal translation mode) related to a degeneracy of the soliton ground state and several finite frequency modes localized near the soliton center. Non zero-frequency modes are expected for the solitons in magnetic dots because the soliton energy depends on the soliton position and soliton radius due to existence of the sample edges. Even though several papers have been already published on the skyrmion dynamical excitations in restricted geometry,\cite{44, 69, gareeva2016magnetic,guslienko2016gyrotropic, mruczkiewicz2016collective} the consensus regarding the classification of the spin eigenmodes over skyrmion background in nanodots has not been reached so far. Also a transition  between the dynamical vortex and skyrmion modes has not been explored. This leaves many questions unanswered, in particular, whether the spin excitation modes corresponding to the two azimuthal SWs existing over a vortex state with clockwise (CW) and counter-clockwise (CCW) sense of propagation can be found in the skyrmion states.

\begin{figure}[!ht]
\includegraphics[width=0.4\textwidth]{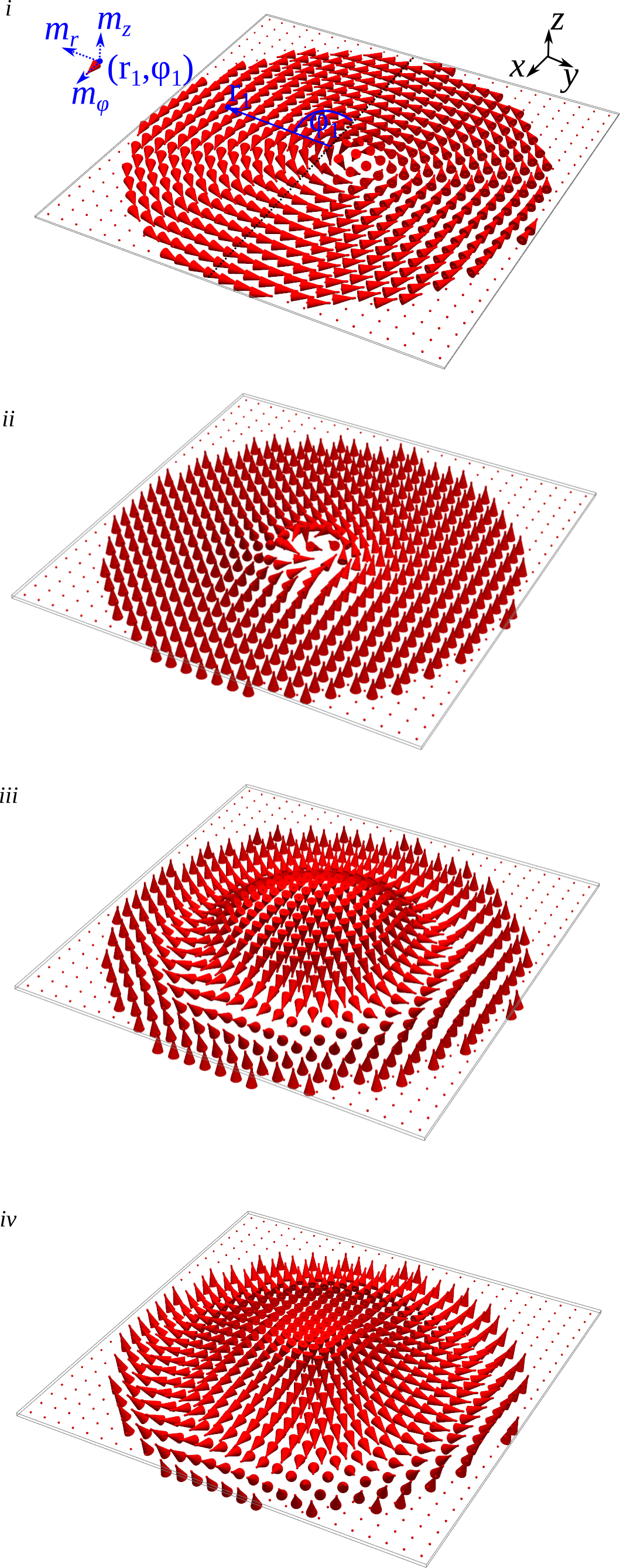}
\caption{(Color online) The four magnetization configurations of the circular ferromagnetic dot investigated in the paper: i) vortex, ii) Bloch-like skyrmion, iii) N\'eel-like skyrmion with strong perpendicular magnetic anisotropy (large value of $Q$) and iv) N\'eel-like skyrmion with small value of $Q$.}
\label{mathematica}
\end{figure}

Analytic approach to the skyrmion dynamics in restricted geometry is related to some model assumptions and simplifications which sometimes are difficult to justify for real systems. For instance, the ideas developed for description of the magnetic bubble domain dynamics were exploited. The skyrmion dynamics in circular dots was considered assuming that all skyrmion excitation modes can be described as oscillation of the circular domain wall shape \cite{makhfudz2012inertia} or that the radial domain wall is very thin (the skyrmion radius is essentially larger than the domain wall width).\cite{gareeva2016magnetic} In this paper we consider spin excitation spectra on the magnetic soliton background in thin circular dots with effective ¨easy plane¨ anisotropy with micromagnetic simulations. To do that we identify a set of parameters that allows for the continuous phase transitions (preserving continuity of the magnetization components)\cite{castel2012perpendicular, taurel2016complete} in transformations between various magnetic soliton stable states, i.e., vortex, Bloch-like skyrmion and N\'eel-like skyrmion (see Fig. \ref{mathematica}). With this we were able to show a continuous mapping of the spin eigenmodes between different magnetization configurations and understand the relation between the excitation modes in different soliton stable states of the nanodots. The results enable to classify the eigenmodes as Bloch-like skyrmion and N\'eel-like skyrmion states and finally to demonstrate existence of the CW and CCW spin wave excitations in these skyrmion state dots.

\section{Model}\label{Sec:Model}
The physical system we consider is a circular ferromagnetic dot of the thickness $t$ and radius $R$. To find the dot spin excitation spectrum the finite difference time domain (FDTD) micromagnetic simulations were performed using mumax$^3$ code.\cite{4899186} We start from the Landau-Lifshitz-Gilbert equation of magnetization $\bf M$ motion with the Gilbert damping parameter in which the time derivative $\frac{\partial {\bf M}({\bf r},t)}{\partial t}$
is defined as the torque ${\bf \tau}$, and equal to:
\begin{equation}
{\bf \tau}= \gamma \frac{1}{1+\alpha^2}\left({\bf M} \times {\bf B_{\mathrm{eff}}}+\alpha \left({\bf M} \times \left({\bf M} \times {\bf B_{\mathrm{eff}}} \right) \right) \right),
\end{equation}
where $\gamma$ is the gyromagnetic ratio, $\alpha$ is a dimensionless damping parameter, and ${\bf B_{\mathrm{eff}}}$ is the effective magnetic field, which includes the external magnetic field ${\bf B_{z}}$,
the magnetostatic demagnetizing field ${\bf B_{\mathrm{demag}}}$,
the isotropic Heisenberg exchange field ${\bf B_{\mathrm{exch}}}$ (the parameter $A_{\text{exch}}$),
the Dzyaloshinskii-Moriya exchange field ${\bf B_{\mathrm{DM}}}$,
and the uniaxial magnetocrystalline anisotropy field ${\bf B_{\mathrm{anis}}}$  (the anisotropy constant $K_u$):
\begin{equation}
{\bf B_{\mathrm{eff}}}={\bf B_{z}}+{\bf B_{\mathrm{demag}}}+{\bf B_{\mathrm{exch}}}+{\bf B_{\mathrm{DM}}}+{\bf B_{\mathrm{anis}}}.
\end{equation}

The skyrmions are stabilized in the dot due to interplay of the isotropic exchange, DMI, uniaxial out-of-plane magnetic anisotropy, and magnetostatic energies assuming zero bias magnetic field. The DMI is implemented as an effective field according to:\cite{bogdanov2001chiral}
\begin{equation}
{\bm B_{D}}=\frac{2D}{M_{s}} \left( \frac{\partial m_{z}}{\partial x},\frac{\partial m_{z}}{\partial y},-\frac{\partial m_{x}}{\partial x}-\frac{\partial m_{y}}{\partial y}  \right)
\end{equation}
that give rise to the energy density:
\begin{equation}
\varepsilon= D \left( m_{z} \left( \nabla \cdot {\bm m} \right) - \left( {\bm m} \cdot \nabla \right)m_{z} \right),
\end{equation}
where ${\bm m} = {\bm M}/M_s$ is the reduced magnetization vector and D is the Dzyaloshinskii-Moriya interface exchange interaction constant.

The simulations consisted of the following steps. We fixed the value of the saturation magnetization $M_s$ and the isotropic exchange $A_{\text{exch}}$. The initial state was assumed in the form of the Bloch skyrmion. This state was relaxed to the lowest local minimum energy state for each set of the parameters ($K_u$, $D$). Then, the stable magnetization configurations were excited with an uniform low amplitude variable magnetic field having a time dependence represented by sinc function with the cut-off frequency $f_{max}=10$ GHz. Such low value of the cut-off frequency was chosen because we were interested to map the low frequency part of the dot spin excitation spectra immediately related to the different dot inhomogeneous magnetization states and continuous transformations
between them. We used in-plane and out-of-plane orientation of the variable field to excite the spin modes of different symmetry. The space and time dependent magnetization components acquired after the field excitation were then transformed to the frequency domain (Fourier transform) to obtain the dispersion of spin modes and spatial distribution of the dynamical components of the selected eigen oscillations of the magnetization vector ${\bm m}$.\cite{mruczkiewicz2016collective}

\begin{figure}[!ht]
\includegraphics[width=0.4\textwidth]{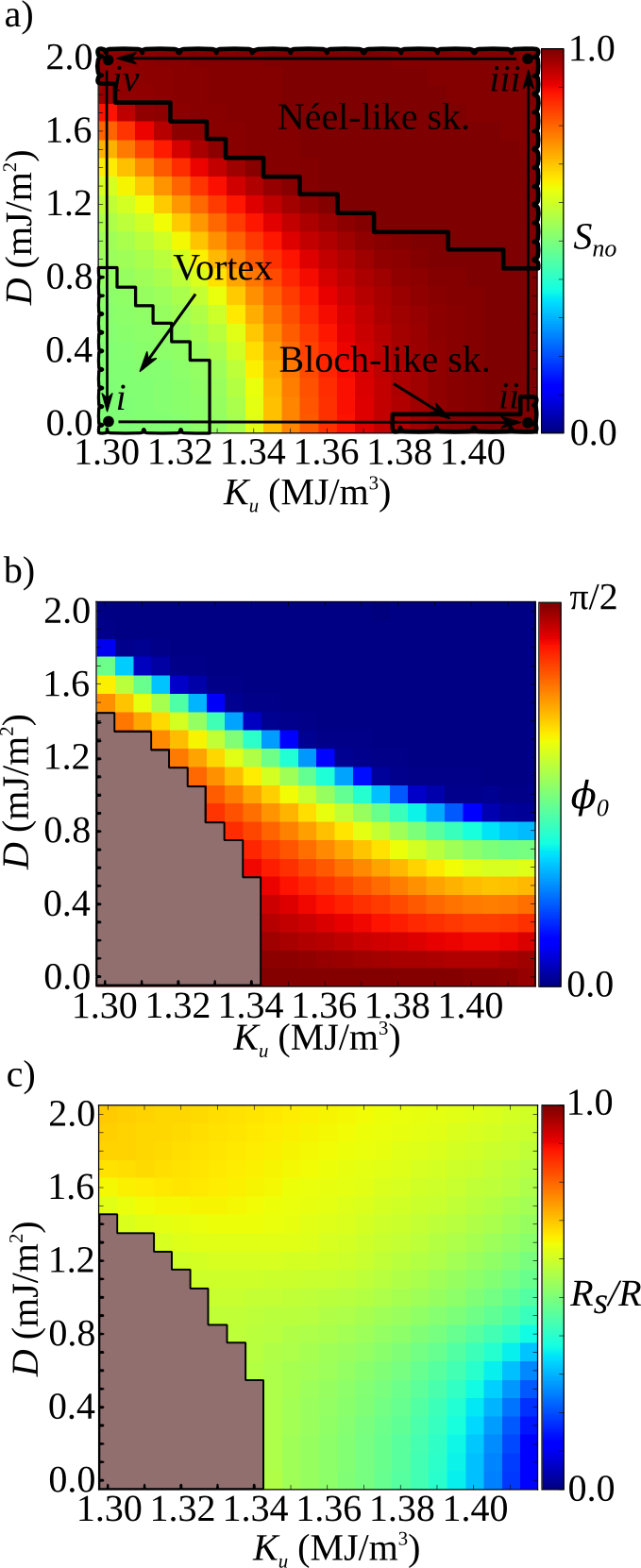}
\caption{(Color online) The properties of the static configuration of the solition in the isolated disk as a function of $D$ and $K_{u}$: a) skyrmion number $S_{no}$, b) skyrmion phase, $\Phi_{0}$ and c) skyrmion size $R_{s}/R$. The points (i-iv) corresponding to magnetic configuration of the Fig. \ref{mathematica} are shown in (a). The arrows indicate the path along which the dispersion relation is plotted in Fig. \ref{disp}. Black line in (a) indicates area of the vortex, Bloch-like and N\'eel-like skyrmion stability.}
\label{static}
\end{figure}

Throughout the paper we use the following material parameters of the ultrathin magnetic circular dot of the radius $R = 125$ nm and thickness $t = 1.4$ nm: saturation magnetization $M_{\text{s}} = 1.5\times 10^6$ A/m, exchange constant $A_{\text{exch}}=3.1 \times10^{-11}$ J/m,
Dzyaloshinskii-Moriya interaction (DMI) constant $D=0$---$2\times 10^{-3}$ J/m$^2$, and out-of-plane magnetic anisotropy constant $K_{u} = 1.30\times 10^6$---$1.415\times 10^6$ J/m$^3$. This set of parameters corresponds to the ultrathin layers of CoFeB-MgO.\cite{nakatani2016electric} The material quality factor $Q=\frac{2 K_{u}}{\mu_{0} M_{\text{s}}^2}$ varies from 0.92 to 1.0. The assumed value of the damping parameter taken into account in the FDTD simulations is $\alpha=0.01$, and it is close to the value of a ultrathin CoFeB film.\cite{Natarajarathinam12,Yu12}

\section{Results and Discussion}
The static soliton stable configurations in the isolated circular dot are shown in Fig. \ref{static}. Fig.~\ref{static}(a) presents the skyrmion number (topological charge):\cite{nagaosa2013topological}
\begin{equation}
S_{no}= \frac{1}{4 \pi} \int \int {\bm m} \cdot \left(\frac{\partial {\bm m}}{\partial x} \times  \frac{\partial {\bm m}}{\partial y}\right) dx dy,
\end{equation}
The topological charge has the characteristic value of 0.5 for the vortex state and 1 for the skyrmion states. However, $S_{no}$ is not exactly equal to 1 for a magnetic skyrmion. It is shown that the $S_{no}$ is a continuous function when the magnetic anisotropy or DMI value are changed in the range of interest defined in the previous section, see Fig. \ref{static} (a). The smooth transition is present from complete vortex (i) to complete Bloch-like (ii) or to N\'eel-like skyrmion (iii) and (iv) (all visualized in Fig.~\ref{mathematica}) with an increase of the anisotropy or DMI value. The Bloch-like skyrmion is stabilized at high values of anisotropy and $D \approx 0$. However, it is a metastable state (a single domain state is the ground state).\cite{nakatani2016electric} The N\'eel-like skyrmion state realized when $D>D_{c}(K_u)$, where the function $D_{c}(K_u)$ is defined by the line following yellow color in Fig. \ref{static} (b). The intermediate states between the vortex and skyrmion are realized for intermediate values of the parameters. In this article we concentrate on the dynamical properties of the nanodot with values of the anisotropy and DMI that links following four points and magnetic configurations indicated in Fig. \ref{mathematica}: i) $D=0$, $K_{u}=1.30 \times 10^6$ J/m$^3$ -- vortex state, ii) $D=0$, $K_{u}=1.415 \times 10^6$ J/m$^3$ -- Bloch-like skyrmion, iii) $D=2\times 10^{-3}$ J/m$^2$, $K_{u}=1.415 \times 10^6$ J/m$^3$ -- N\'eel-like skyrmion with high quality factor $Q$ and iv) $D=2\times 10^{-3}$ J/m$^2$, $K_{u}=1.30 \times 10^6$ J/m$^3$ -- N\'eel-like skyrmion with low $Q$ ($Q=0.92$).

The magnetization vector dependence on the coordinates can be expressed in the polar coordinate system ${\bm m}(r,\phi)$ with the origin in the dot center, see Fig. \ref{mathematica} (i). Then, the static and dynamical magnetization are expressed via their polar coordinate components,  $m_{\phi}$, $m_{r}$, $m_{z}$ and $\delta m_{\phi}$, $\delta m_{r}$, $\delta m_{z}$, respectively. We express the magnetization components via the magnetization spherical angles, $\Theta$ and $\Phi$.  For radially symmetric static magnetization configurations $\Theta= \Theta(r)$ and $\Phi = \Phi_{0} + \phi$. The type of the skyrmion (Bloch-like or N\'eel-like) are distinguished with a function $\Phi_{0}$, skyrmion phase. It takes a value $ +/- \frac{\pi}{2}$ for the magnetic vortex or complete Bloch-like skyrmion (magnetization is aligned along the azimuthal direction at the vortex/skyrmion edge) and $0, \pi$ for the complete N\'eel-like skyrmion (magnetization is along the radial direction). The different values of $\Phi_{0}$ describe the vortex/skyrmion chirality and one of them should be chosen for the defined sign of the DMI parameter $D$.  However, for some intermediate magnetization states, $\Phi_{0}$ is a function of the position $r$. Therefore, we define $\Phi_{0}$ from the orientation of the magnetization at the value of the radial coordinate equal to the skyrmion radius, $R_{s}$ (defined by the condition $m_{z}(r=R_{s},\phi)=0$):
\begin{equation}
\Phi_{0}= \arcsin{\frac{m_{\phi}}{\sqrt{m_{r}^2+m_{\phi}^2}}}.
\end{equation}
The function $\Phi_{0}$ over $(D,K_{u})$ plane is presented in Fig.~\ref{static} (b). The $\Phi_{0}$ is defined only when one point $m_{z}=0$ can be found along the dot radius. The general tendency is that the low value of DMI favors Bloch-like skyrmion (ii) and the high value of DMI is necessary to stabilize the N\'eel-like skyrmion (iii) and (iv). The vortex is stable when both $(D, K_{u})$ values are small.

Fig.~\ref{static} (c) shows the dependence of the skyrmion size $R_{s}$  normalized to the dot radius $R$. Similarly to the $\Phi_{0}$, $R_{s}$ is defined only when the point $m_{z}=0$ can be found. The $R_{s}$ is also calculated for the incomplete skyrmions having the topological charge smaller than 1, where still the point $m_{z}=0$ can be found along the dot radial coordinate. The size of the skyrmion increases with the increase of the DMI strength and decreases with increase of the magnetic anisotropy constant $K_{u}$.\cite{moreau2016additive,7061384} The Bloch-like skyrmion becomes unstable ($R_{s}$ goes to zero) with respect to transition to the perpendicular single-domain state at $Q$ approaching 1 and small $D<D_{c}(K_u)=0.6\times 10^{-3}$ J/m$^3$.

The analysis of Fig.~\ref{static} allows us to choose the path for study of the soliton dynamical excitations where continuous transitions (second-order phase transition) happen: between the vortex and Bloch-like skyrmion along path (i) $\rightarrow$ (ii), between the Bloch-like skyrmion and N\'eel-like skyrmion with high $Q$ along path (ii) $\rightarrow$ (iii), between high and low $Q$ N\'eel-like skyrmions along the path (iii) $\rightarrow$ (iv), and finally between the N\'eel-like skyrmion and vortex state along  path (iii) $\rightarrow$ (iv). These paths are indicated in Fig. \ref{static} (a) with black straight arrows.

\begin{figure}[!ht]
\includegraphics[width=0.45\textwidth]{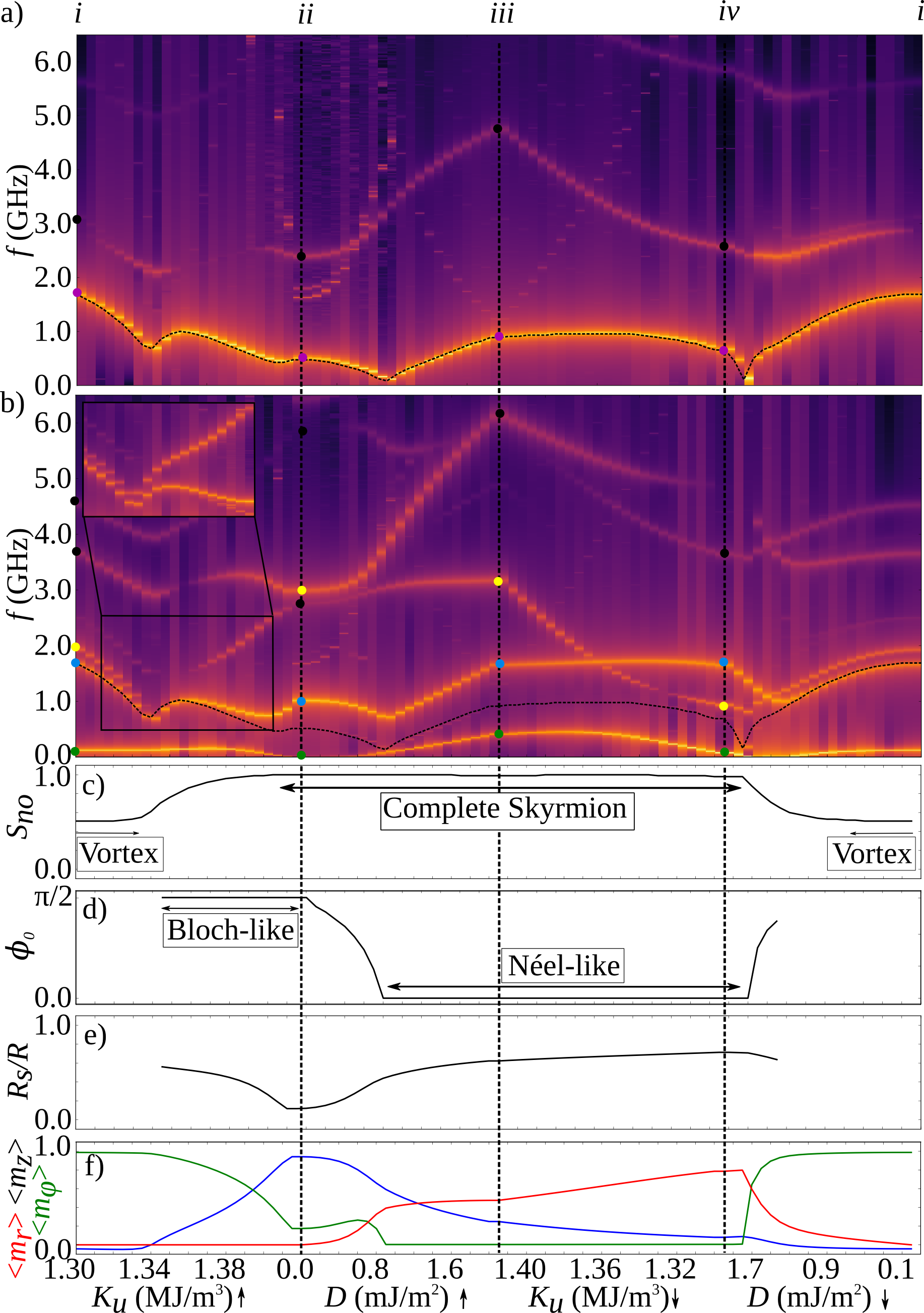}
\caption{(Color online) The dispersion relation plotted along the path presented in Fig. \ref{static}(a) between the four (i-iv) magnetic configurations presented in Fig. \ref{mathematica}. The figure presents the  dispersion obtained with the spatially uniform microwave magnetic field excitation of the sinc type in time with a) out-of-plane dynamic component and b) in-plane magnetization component. The static properties of the soliton are presented in c) skyrmion number $S_{no}$, d) skyrmion phase, $\Phi_{0}$, e) skyrmion size $R_{s}/R$ and f) the averaged magnetization components, $<m_{r}>$, $<m_{\phi}>$ and $<m_{z}>$. }
\label{disp}
\end{figure}

In the dynamical simulations, there are spin eigenmodes keeping the radial symmetry of the soliton static state (radially symmetric or breathing modes) and the eigenmodes with broken radial symmetry (azimuthal modes rotating in the clockwise (CW) and counter-clockwise (CCW) directions). The radially symmetric modes have no net in-plane magnetization, and therefore, can be excited only by out-of-plane variable magnetic field. The azimuthal modes including gyrotropic ones can be excited by an in-plane variable magnetic field.
The calculated frequencies of low-lying spin excitations along the defined above paths are presented in Fig. \ref{disp} (a) for the out-of-plane excitation magnetic field and in Fig. \ref{disp} (b) for the in-plane excitation magnetic field. The static properties presented in Fig. \ref{static} are also plotted as a function of the dot magnetic parameters in Figs. \ref{disp} (c-f). The regions where soliton is in complete vortex, Bloch-like, N\'eel-like skyrmions or in transition state can be clearly differentiated and are indicated in Fig.~\ref{disp} (c) and (d), and also in Fig.~\ref{static}(a). Fig.~\ref{disp}(f) demonstrate the second-order phase transitions, i.e., continuous transitions of the average magnetization components (its perpendicular $<m_{z}>$, radial $<m_{r}>$ and azimuthal $<m_{\phi}>$ components) along the paths.

\begin{figure*}

 \center

  \includegraphics[width=0.75\textwidth]{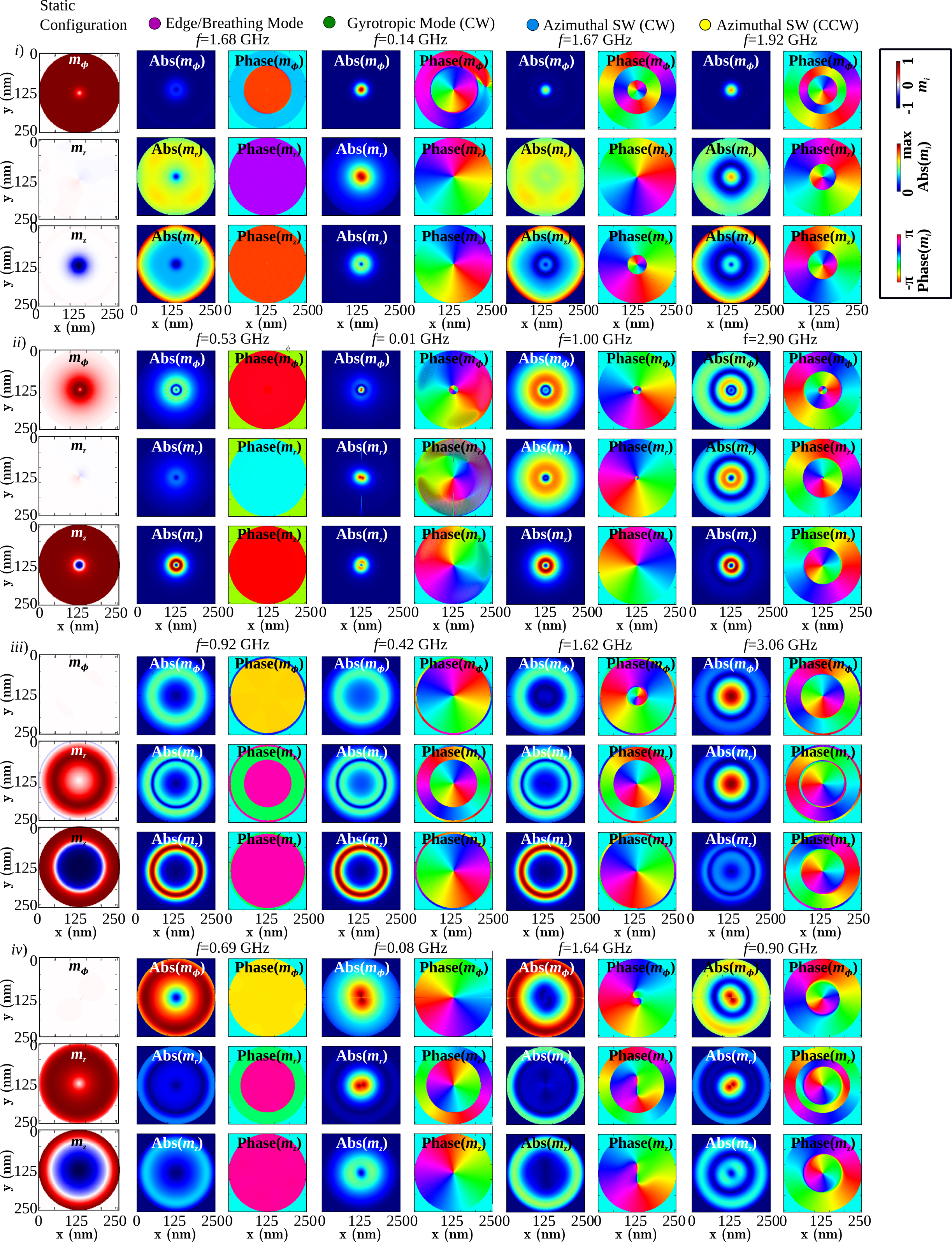}

  \caption{(Color online) The static (1st column on the left) and dynamic magnetization components, $\delta m_{\phi}$, $\delta m_{r}$ and $\delta m_{z}$ of the isolated disk in soliton magnetic configurations in the vortex (i), Bloch-like (ii) and N\'eel-like (iii),  and (iv) skyrmions. The magnetic configuration are presented on the Fig. \ref{mathematica}, the frequency dispersion is indicated as grid vertical line on the Fig. \ref{disp} (a) and (b).}

  \label{prof}

\end{figure*}

Two main modes can be found in the excitation spectrum shown in Fig.~\ref{disp}(a) for the external microwave field perpendicular to the dot plane. The lowest frequency mode (purple dot and black dashed line) corresponds to the so called breathing skyrmion modes.\cite{44} The higher frequency mode (marked by black dot) is a high order quantized mode of the same type of the radially symmetric excitations. Continuous transition of the mode along the path allows to find corresponding modes in the vortex state, in the Bloch-like and N\'eel-like skyrmions. The spatial profiles of the dynamic magnetization of the lowest frequency breathing mode for the four magnetic configurations (i-iv) are plotted in Fig.~\ref{prof} (2nd column). For the vortex magnetic configuration, $K_{u}=1.3 \times 10^6$ J/m$^3$  and $D=0$ mJ/m$^2$ (i) this mode can be characterized as a almost uniform radial mode. The SW amplitude is connected mainly with radial magnetization component, that for static configuration in a vortex state is close to 0 (see the 1st column in Fig.~\ref{prof}), and it oscillates in phase in a whole nanodot. With increase of the anisotropy the magnetic configuration is transformed to the Bloch-like skyrmion (point (ii)) and the character of the mode changes. The out-of-plane dynamic component is localized near the edge of skyrmion (near the skyrmion radius $r=R_{s}$),  forms a ring around the skyrmion edge and the magnetization oscillations are in phase.  That is characteristic for the lowest breathing mode. Although the size of skyrmion is smallest among the considered configurations (Fig.~\ref{static}(e)) the frequency of the breathing mode is quite low. With increasing $D$, its frequency still decreases up to transformation of the soliton into the N\'eel-like skyrmion, whereas the skyrmion radius increases. The breathing character of the mode and the area of the mode localization are preserved for the N\'eel-like skyrmion with high $Q$ factor, $K_{u}=1.415 \times 10^6$ J/m$^3$  and $D=2.0$ mJ/m$^2$ (iii), as well. Whereas for the  N\'eel-like skyrmion with low $Q$ factor, at the point (iv) the amplitudes of the $z$ and $\phi$ dynamical magnetization components are localized near the dot edge and the largest value of the static $z$-component amplitude is connected with the largest $\delta m_{\phi}$ component. This change might be attributed to large size of the skyrmion and related edge effects or the large domain wall width related to the low value of $Q$ factor.\cite{kisielewski2003drastic, virot2012theory} Interestingly, the frequency of the breathing mode has a local minimum at every transformation of the magnetization configuration (between vortex and Bloch-like skyrmion, Bloch-like and N\'eel-like skyrmion, and N\'eel-like skyrmion and the vortex), while it is weakly dependent on the changes of $Q$ for the N\'eel-like skyrmion.

Fig. \ref{disp} (b) presents the soliton eigenfrequencies when the excitation magnetic field is in-plane of the dot. Due to the symmetry of the breathing mode, it cannot be excited effectively with such field. Instead, other modes are visible in the excitation spectrum. Here we concentrate on the modes that correspond to gyrotropic  mode and CW, CCW azimuthal SW modes in the vortex magnetic configuration indicated with green, blue and yellow dots, respectively.

The lowest frequency mode is a gyrotropic excitation directly related to the soliton topological charge. This mode is a precession of the vortex/skyrmion core around its equilibrium position in the dot center (CW for the given soliton core polarization $p$=-1). It is characterized with the out-of-plane dynamic magnetization component localized near the edge of the skyrmion and amplitude that forms a ring around the skyrmion edge (see Fig. \ref{prof}, the column in the middle). The oscillation is not in phase around the dot, but changes continuously form $-\pi$ to $\pi$. This character is preserved for all configurations apart from the N\'eel-like skyrmion with low $Q$ factor, where due to large size of the skyrmion, the nanodot edge effects or the wide domain wall width provide changes in the profile of this mode. The in-plane dynamical magnetization components are also concentrated near the skyrmion radius except the case of N\'eel-like skyrmion (iv), where they are localized at the dot center. The CW gyrotropic mode can be described by the azimuthal index $m=-1$. The gyrotropic mode posses a finite frequency in the sub-GHz range for the whole studied range of parameters with two minima and close to zero frequency for Bloch-like skyrmion (around ii) and N\'eel-like skyrmion  with low $Q$ factor (near iv). The vanishing frequency is a characteristic property of the gyrotropic mode signaling about a border of the soliton stability. It was also observed near the transition of the vortex to the saturated state of the dots, when the gyrotropic mode transforms into the quasi-uniform Kittel mode.\cite{taurel2016complete} The minimum of frequency corresponds to two magnetization configurations with smallest (ii) and largest (iv) skyrmion radius, the states closest to the uniform state.

Two remaining excitation modes in the vortex state are the modes corresponding to the CW ($m=-1$) and CCW ($m=+1$) SW azimuthal modes (see profiles in Fig.~\ref{prof} in the last two columns). If the vortex core is neglected they have no radial nodes and can be described by the radial index $n=0$. In a vortex state (i) the degeneracy of the CW and CCW SWs is lifted due to the dynamic hybridization of the SWs with the gyrotropic mode.\cite{guslienko2008dynamic} This hybridization is especially strong for the $m=-1$ SW mode and small $R$ resulting in formation of a radial node near the vortex core edge. Therefore, the high frequency mode of the doublet is described using the indices $m=-1, n=1$. The classification of the low-lying SW modes by the indices $m=+1, n=0$ (CCW) and $m=-1, n=1$ (CW) is applicable also to the Bloch- and Neel-skyrmions. The mode frequencies decrease with increasing $K_u$ when the magnetic stable state is vortex. During transition from vortex to Bloch-like skyrmion, the spatial distribution of CCW SW mode changes. The localization is transferred from the edge to center and at certain values of anisotropy the mode can not be effectively excited with in-plane field. To check if the separation of the CW and CCW modes preserves in the transition point we have repeated simulation but with the point excitation (100 nm circle localized in the center of the nanodot). Indeed, these modes are split as is shown in the inset in Fig.~\ref{disp} in the transition point. With further increase of the anisotropy the Bloch-like skyrmion forms and the frequency separation between CW and CCW modes increases significantly. It is related to the localization of these modes in the skyrmion states, which is different from one in the vortex state where the vortex core occupies small area of the dot.  In the skyrmion states we can distinguish azimuthal modes as center localized or edge localized SWs.  With transformation of the magnetization configuration between different skyrmions this property is preserved and mostly pronounced among the skyrmion states in the state (iii) where strong localization of the azimuthal SW is present at a skyrmion core (5th column in Fig.~\ref{prof}) or skyrmion edge (4th column). Nevertheless, the origin of this frequency splitting is not clear at this moment and gyrotropic mode can contribute to this. \cite{gareeva2016magnetic}

In the case of N\'eel-like skyrmion configuration with high $Q$ factor (iii), the lower frequency azimuthal SW could be also interpreted as CCW gyrotropic mode\cite{mochizuki2012spin} and in fact, there is a big resemblance of the dynamic components of magnetization. Nevertheless, the origin of this mode is due to azimuthal SW excitation over the skyrmion background and the difference between the gyrotropic mode and lowest azimuthal SW mode is pronounced when the size of the skyrmion is much smaller than the dot size, the localization of the in-plane magnetization components of the azimuthal SW mode are spread within the dot and the in-plane magnetization components of the gyrotropic mode are localized near the skyrmion edge (ii). Other difference is the link of the gyrotropic mode sense of rotation with the sign of the topological charge or gyrovector ($p=+1/-1$). With respect of the CW and CCW SW mode frequencies splitting, interesting is also the variation of these modes with decreasing $K_u$ in the N\'eel-like skyrmion (on the path between iii and iv), where the frequency order of these modes changes, without any signature of their interaction at their frequencies crossing. In the Neel-skyrmion state (iv) the spin excitation spectrum is consequence of the CW gyrotropic ($m=-1$), CW SW ($m=-1, n=1$) and CCW SW ($m=+1, n=0$) modes. Whereas, the consequence for the vortex, Bloch- (ii) and Neel-skyrmion states is the following: CW gyrotropic ($m=-1$), CCW SW ($m=+1, n=0$) and CW SW ($m=-1, n=1$) modes. Other peculiarity of the Neel-skyrmion (iv) state is strong localization of the high frequency CCW SW mode near the dot edge.

We note also that there is a correspondence between azimuthal CCW SW mode $m=+1, n=0$ (blue dots) and uniform radial mode $m=0, n=0$ (purple dots) in the dynamical magnetization spatial distribution and the frequency dependence on the parameters $D, K_u$ for all magnetic configurations considered. This similarity is indicated in Fig. \ref{disp}(a) and (b) by dashed black line.

\section{Summary}\label{Sec:Conclusions}

We have determined the spin excitation spectra of a circular magnetic dot in four topologically non-trivial magnetization states, and showed the continuous transitions between them. Using spin-wave eigenmode mapping we have elucidated the origin of the excitations on the skyrmion background and proposed a spin mode classification similar to that developed for the spin excitations in the magnetic vortex state dots (with $K_u=0$) based on the azimuthal and radial mode indices $(m, n)$. The excitation spectrum on the vortex background with large $K_u$ is similar to that of the vortex state in a soft magnetic dot in that it has a low-frequency gyrotropic mode, an $m=+1/-1$ azimuthal mode doublet, and a radial mode ($m=0, n=0$) with approximately the same frequencies. In the skyrmion state the radial mode frequency is essentially lower and the frequency splitting between the azimuthal ($m=+1$ and $m=-1$) spin-wave modes essentially larger than ones in a vortex state dot.


\section*{Acknowledgements}

One of the co-authors (K.G.) thanks to B.A. Ivanov for discussions of the magnetic soliton dynamics. The project is  financed from the SASPRO Programme. The research has received  funding from the People Programme (Marie  Curie  Actions) European Union's Seventh Framework Programme under REA grant agreement No. 609427 (Project WEST: 1244/02/01). Research has been further co-funded by the Slovak Academy of Sciences and the European Union Horizon 2020 Research and Innovation Programme under Marie Sklodowska-Curie Grant Agreement No.~644348 (MagIC). K.G. acknowledges support by IKERBASQUE (the Basque Foundation for Science), the Spanish MINECO grant MAT2013-47078-C2-1-P. The numerical calculations were partially performed at the Poznan Supercomputing and Networking Center (Grant No.~209).

\bibliographystyle{apsrev4-1}
\bibliography{books}

\begin{thebibliography}{36}%
\makeatletter
\providecommand \@ifxundefined [1]{%
 \@ifx{#1\undefined}
}%
\providecommand \@ifnum [1]{%
 \ifnum #1\expandafter \@firstoftwo
 \else \expandafter \@secondoftwo
 \fi
}%
\providecommand \@ifx [1]{%
 \ifx #1\expandafter \@firstoftwo
 \else \expandafter \@secondoftwo
 \fi
}%
\providecommand \natexlab [1]{#1}%
\providecommand \enquote  [1]{``#1''}%
\providecommand \bibnamefont  [1]{#1}%
\providecommand \bibfnamefont [1]{#1}%
\providecommand \citenamefont [1]{#1}%
\providecommand \href@noop [0]{\@secondoftwo}%
\providecommand \href [0]{\begingroup \@sanitize@url \@href}%
\providecommand \@href[1]{\@@startlink{#1}\@@href}%
\providecommand \@@href[1]{\endgroup#1\@@endlink}%
\providecommand \@sanitize@url [0]{\catcode `\\12\catcode `\$12\catcode
  `\&12\catcode `\#12\catcode `\^12\catcode `\_12\catcode `\%12\relax}%
\providecommand \@@startlink[1]{}%
\providecommand \@@endlink[0]{}%
\providecommand \url  [0]{\begingroup\@sanitize@url \@url }%
\providecommand \@url [1]{\endgroup\@href {#1}{\urlprefix }}%
\providecommand \urlprefix  [0]{URL }%
\providecommand \Eprint [0]{\href }%
\providecommand \doibase [0]{http://dx.doi.org/}%
\providecommand \selectlanguage [0]{\@gobble}%
\providecommand \bibinfo  [0]{\@secondoftwo}%
\providecommand \bibfield  [0]{\@secondoftwo}%
\providecommand \translation [1]{[#1]}%
\providecommand \BibitemOpen [0]{}%
\providecommand \bibitemStop [0]{}%
\providecommand \bibitemNoStop [0]{.\EOS\space}%
\providecommand \EOS [0]{\spacefactor3000\relax}%
\providecommand \BibitemShut  [1]{\csname bibitem#1\endcsname}%
\let\auto@bib@innerbib\@empty
\bibitem [{\citenamefont {Kosevich}\ \emph {et~al.}(1990)\citenamefont
  {Kosevich}, \citenamefont {Ivanov},\ and\ \citenamefont
  {Kovalev}}]{kosevich1990magnetic}%
  \BibitemOpen
  \bibfield  {author} {\bibinfo {author} {\bibfnamefont {A.~M.}\ \bibnamefont
  {Kosevich}}, \bibinfo {author} {\bibfnamefont {B.}~\bibnamefont {Ivanov}}, \
  and\ \bibinfo {author} {\bibfnamefont {A.}~\bibnamefont {Kovalev}},\
  }\href@noop {} {\bibfield  {journal} {\bibinfo  {journal} {Physics Reports}\
  }\textbf {\bibinfo {volume} {194}},\ \bibinfo {pages} {117} (\bibinfo {year}
  {1990})}\BibitemShut {NoStop}%
\bibitem [{\citenamefont {Shinjo}\ \emph {et~al.}(2000)\citenamefont {Shinjo},
  \citenamefont {Okuno}, \citenamefont {Hassdorf}, \citenamefont {Shigeto},\
  and\ \citenamefont {Ono}}]{shinjo2000magnetic}%
  \BibitemOpen
  \bibfield  {author} {\bibinfo {author} {\bibfnamefont {T.}~\bibnamefont
  {Shinjo}}, \bibinfo {author} {\bibfnamefont {T.}~\bibnamefont {Okuno}},
  \bibinfo {author} {\bibfnamefont {R.}~\bibnamefont {Hassdorf}}, \bibinfo
  {author} {\bibfnamefont {K.}~\bibnamefont {Shigeto}}, \ and\ \bibinfo
  {author} {\bibfnamefont {T.}~\bibnamefont {Ono}},\ }\href@noop {} {\bibfield
  {journal} {\bibinfo  {journal} {Science}\ }\textbf {\bibinfo {volume}
  {289}},\ \bibinfo {pages} {930} (\bibinfo {year} {2000})}\BibitemShut
  {NoStop}%
\bibitem [{\citenamefont {Guslienko}\ \emph {et~al.}(2002)\citenamefont
  {Guslienko}, \citenamefont {Ivanov}, \citenamefont {Novosad}, \citenamefont
  {Otani}, \citenamefont {Shima},\ and\ \citenamefont
  {Fukamichi}}]{guslienko2002eigenfrequencies}%
  \BibitemOpen
  \bibfield  {author} {\bibinfo {author} {\bibfnamefont {K.~Y.}\ \bibnamefont
  {Guslienko}}, \bibinfo {author} {\bibfnamefont {B.}~\bibnamefont {Ivanov}},
  \bibinfo {author} {\bibfnamefont {V.}~\bibnamefont {Novosad}}, \bibinfo
  {author} {\bibfnamefont {Y.}~\bibnamefont {Otani}}, \bibinfo {author}
  {\bibfnamefont {H.}~\bibnamefont {Shima}}, \ and\ \bibinfo {author}
  {\bibfnamefont {K.}~\bibnamefont {Fukamichi}},\ }\href@noop {} {\bibfield
  {journal} {\bibinfo  {journal} {Journal of Applied Physics}\ }\textbf
  {\bibinfo {volume} {91}},\ \bibinfo {pages} {8037} (\bibinfo {year}
  {2002})}\BibitemShut {NoStop}%
\bibitem [{\citenamefont {Guslienko}\ \emph {et~al.}(2006)\citenamefont
  {Guslienko}, \citenamefont {Han}, \citenamefont {Keavney}, \citenamefont
  {Divan},\ and\ \citenamefont {Bader}}]{guslienko2006magnetic}%
  \BibitemOpen
  \bibfield  {author} {\bibinfo {author} {\bibfnamefont {K.~Y.}\ \bibnamefont
  {Guslienko}}, \bibinfo {author} {\bibfnamefont {X.}~\bibnamefont {Han}},
  \bibinfo {author} {\bibfnamefont {D.}~\bibnamefont {Keavney}}, \bibinfo
  {author} {\bibfnamefont {R.}~\bibnamefont {Divan}}, \ and\ \bibinfo {author}
  {\bibfnamefont {S.}~\bibnamefont {Bader}},\ }\href@noop {} {\bibfield
  {journal} {\bibinfo  {journal} {Physical Review Letters}\ }\textbf {\bibinfo
  {volume} {96}},\ \bibinfo {pages} {067205} (\bibinfo {year}
  {2006})}\BibitemShut {NoStop}%
\bibitem [{\citenamefont {Park}\ and\ \citenamefont {Crowell}(2005)}]{112}%
  \BibitemOpen
  \bibfield  {author} {\bibinfo {author} {\bibfnamefont {J.}~\bibnamefont
  {Park}}\ and\ \bibinfo {author} {\bibfnamefont {P.}~\bibnamefont {Crowell}},\
  }\href@noop {} {\bibfield  {journal} {\bibinfo  {journal} {Physical Review
  Letters}\ }\textbf {\bibinfo {volume} {95}},\ \bibinfo {pages} {167201}
  (\bibinfo {year} {2005})}\BibitemShut {NoStop}%
\bibitem [{\citenamefont {Mamica}\ \emph {et~al.}(2013)\citenamefont {Mamica},
  \citenamefont {L{\'e}vy},\ and\ \citenamefont
  {Krawczyk}}]{mamica2013effects}%
  \BibitemOpen
  \bibfield  {author} {\bibinfo {author} {\bibfnamefont {S.}~\bibnamefont
  {Mamica}}, \bibinfo {author} {\bibfnamefont {J.~S.}\ \bibnamefont
  {L{\'e}vy}}, \ and\ \bibinfo {author} {\bibfnamefont {M.}~\bibnamefont
  {Krawczyk}},\ }\href@noop {} {\bibfield  {journal} {\bibinfo  {journal}
  {Journal of Physics D: Applied Physics}\ }\textbf {\bibinfo {volume} {47}},\
  \bibinfo {pages} {015003} (\bibinfo {year} {2013})}\BibitemShut {NoStop}%
\bibitem [{\citenamefont {Guslienko}\ \emph {et~al.}(2001)\citenamefont
  {Guslienko}, \citenamefont {Novosad}, \citenamefont {Otani}, \citenamefont
  {Shima},\ and\ \citenamefont {Fukamichi}}]{guslienko2001magnetization}%
  \BibitemOpen
  \bibfield  {author} {\bibinfo {author} {\bibfnamefont {K.~Y.}\ \bibnamefont
  {Guslienko}}, \bibinfo {author} {\bibfnamefont {V.}~\bibnamefont {Novosad}},
  \bibinfo {author} {\bibfnamefont {Y.}~\bibnamefont {Otani}}, \bibinfo
  {author} {\bibfnamefont {H.}~\bibnamefont {Shima}}, \ and\ \bibinfo {author}
  {\bibfnamefont {K.}~\bibnamefont {Fukamichi}},\ }\href@noop {} {\bibfield
  {journal} {\bibinfo  {journal} {Physical Review B}\ }\textbf {\bibinfo
  {volume} {65}},\ \bibinfo {pages} {024414} (\bibinfo {year}
  {2001})}\BibitemShut {NoStop}%
\bibitem [{\citenamefont {Guslienko}\ \emph {et~al.}(2008)\citenamefont
  {Guslienko}, \citenamefont {Slavin}, \citenamefont {Tiberkevich},\ and\
  \citenamefont {Kim}}]{guslienko2008dynamic}%
  \BibitemOpen
  \bibfield  {author} {\bibinfo {author} {\bibfnamefont {K.~Y.}\ \bibnamefont
  {Guslienko}}, \bibinfo {author} {\bibfnamefont {A.~N.}\ \bibnamefont
  {Slavin}}, \bibinfo {author} {\bibfnamefont {V.}~\bibnamefont {Tiberkevich}},
  \ and\ \bibinfo {author} {\bibfnamefont {S.-K.}\ \bibnamefont {Kim}},\
  }\href@noop {} {\bibfield  {journal} {\bibinfo  {journal} {Physical Review
  Letters}\ }\textbf {\bibinfo {volume} {101}},\ \bibinfo {pages} {247203}
  (\bibinfo {year} {2008})}\BibitemShut {NoStop}%
\bibitem [{\citenamefont {Pigeau}\ \emph {et~al.}(2010)\citenamefont {Pigeau},
  \citenamefont {De~Loubens}, \citenamefont {Klein}, \citenamefont {Riegler},
  \citenamefont {Lochner}, \citenamefont {Schmidt}, \citenamefont {Molenkamp},
  \citenamefont {Tiberkevich},\ and\ \citenamefont {Slavin}}]{114}%
  \BibitemOpen
  \bibfield  {author} {\bibinfo {author} {\bibfnamefont {B.}~\bibnamefont
  {Pigeau}}, \bibinfo {author} {\bibfnamefont {G.}~\bibnamefont {De~Loubens}},
  \bibinfo {author} {\bibfnamefont {O.}~\bibnamefont {Klein}}, \bibinfo
  {author} {\bibfnamefont {A.}~\bibnamefont {Riegler}}, \bibinfo {author}
  {\bibfnamefont {F.}~\bibnamefont {Lochner}}, \bibinfo {author} {\bibfnamefont
  {G.}~\bibnamefont {Schmidt}}, \bibinfo {author} {\bibfnamefont
  {L.}~\bibnamefont {Molenkamp}}, \bibinfo {author} {\bibfnamefont
  {V.}~\bibnamefont {Tiberkevich}}, \ and\ \bibinfo {author} {\bibfnamefont
  {A.}~\bibnamefont {Slavin}},\ }\href@noop {} {\bibfield  {journal} {\bibinfo
  {journal} {Applied Physics Letters}\ }\textbf {\bibinfo {volume} {96}},\
  \bibinfo {pages} {132506} (\bibinfo {year} {2010})}\BibitemShut {NoStop}%
\bibitem [{\citenamefont {Mochizuki}(2012{\natexlab{a}})}]{38}%
  \BibitemOpen
  \bibfield  {author} {\bibinfo {author} {\bibfnamefont {M.}~\bibnamefont
  {Mochizuki}},\ }\href@noop {} {\bibfield  {journal} {\bibinfo  {journal}
  {Physical Review Letters}\ }\textbf {\bibinfo {volume} {108}},\ \bibinfo
  {pages} {017601} (\bibinfo {year} {2012}{\natexlab{a}})}\BibitemShut
  {NoStop}%
\bibitem [{\citenamefont {Okamura}\ \emph {et~al.}(2013)\citenamefont
  {Okamura}, \citenamefont {Kagawa}, \citenamefont {Mochizuki}, \citenamefont
  {Kubota}, \citenamefont {Seki}, \citenamefont {Ishiwata}, \citenamefont
  {Kawasaki}, \citenamefont {Onose},\ and\ \citenamefont {Tokura}}]{57}%
  \BibitemOpen
  \bibfield  {author} {\bibinfo {author} {\bibfnamefont {Y.}~\bibnamefont
  {Okamura}}, \bibinfo {author} {\bibfnamefont {F.}~\bibnamefont {Kagawa}},
  \bibinfo {author} {\bibfnamefont {M.}~\bibnamefont {Mochizuki}}, \bibinfo
  {author} {\bibfnamefont {M.}~\bibnamefont {Kubota}}, \bibinfo {author}
  {\bibfnamefont {S.}~\bibnamefont {Seki}}, \bibinfo {author} {\bibfnamefont
  {S.}~\bibnamefont {Ishiwata}}, \bibinfo {author} {\bibfnamefont
  {M.}~\bibnamefont {Kawasaki}}, \bibinfo {author} {\bibfnamefont
  {Y.}~\bibnamefont {Onose}}, \ and\ \bibinfo {author} {\bibfnamefont
  {Y.}~\bibnamefont {Tokura}},\ }\href@noop {} {\bibfield  {journal} {\bibinfo
  {journal} {Nature Communications}\ }\textbf {\bibinfo {volume} {4}},\
  \bibinfo {pages} {2391} (\bibinfo {year} {2013})}\BibitemShut {NoStop}%
\bibitem [{\citenamefont {Schwarze}\ \emph {et~al.}(2015)\citenamefont
  {Schwarze}, \citenamefont {Waizner}, \citenamefont {Garst}, \citenamefont
  {Bauer}, \citenamefont {Stasinopoulos}, \citenamefont {Berger}, \citenamefont
  {Pfleiderer},\ and\ \citenamefont {Grundler}}]{schwarze2015universal}%
  \BibitemOpen
  \bibfield  {author} {\bibinfo {author} {\bibfnamefont {T.}~\bibnamefont
  {Schwarze}}, \bibinfo {author} {\bibfnamefont {J.}~\bibnamefont {Waizner}},
  \bibinfo {author} {\bibfnamefont {M.}~\bibnamefont {Garst}}, \bibinfo
  {author} {\bibfnamefont {A.}~\bibnamefont {Bauer}}, \bibinfo {author}
  {\bibfnamefont {I.}~\bibnamefont {Stasinopoulos}}, \bibinfo {author}
  {\bibfnamefont {H.}~\bibnamefont {Berger}}, \bibinfo {author} {\bibfnamefont
  {C.}~\bibnamefont {Pfleiderer}}, \ and\ \bibinfo {author} {\bibfnamefont
  {D.}~\bibnamefont {Grundler}},\ }\href@noop {} {\bibfield  {journal}
  {\bibinfo  {journal} {Nature Materials}\ }\textbf {\bibinfo {volume} {14}},\
  \bibinfo {pages} {478} (\bibinfo {year} {2015})}\BibitemShut {NoStop}%
\bibitem [{\citenamefont {Moreau-Luchaire}\ \emph {et~al.}(2016)\citenamefont
  {Moreau-Luchaire}, \citenamefont {Moutafis}, \citenamefont {Reyren},
  \citenamefont {Sampaio}, \citenamefont {Vaz}, \citenamefont {Van~Horne},
  \citenamefont {Bouzehouane}, \citenamefont {Garcia}, \citenamefont
  {Deranlot}, \citenamefont {Warnicke} \emph {et~al.}}]{moreau2016additive}%
  \BibitemOpen
  \bibfield  {author} {\bibinfo {author} {\bibfnamefont {C.}~\bibnamefont
  {Moreau-Luchaire}}, \bibinfo {author} {\bibfnamefont {C.}~\bibnamefont
  {Moutafis}}, \bibinfo {author} {\bibfnamefont {N.}~\bibnamefont {Reyren}},
  \bibinfo {author} {\bibfnamefont {J.}~\bibnamefont {Sampaio}}, \bibinfo
  {author} {\bibfnamefont {C.}~\bibnamefont {Vaz}}, \bibinfo {author}
  {\bibfnamefont {N.}~\bibnamefont {Van~Horne}}, \bibinfo {author}
  {\bibfnamefont {K.}~\bibnamefont {Bouzehouane}}, \bibinfo {author}
  {\bibfnamefont {K.}~\bibnamefont {Garcia}}, \bibinfo {author} {\bibfnamefont
  {C.}~\bibnamefont {Deranlot}}, \bibinfo {author} {\bibfnamefont
  {P.}~\bibnamefont {Warnicke}},  \emph {et~al.},\ }\href@noop {} {\bibfield
  {journal} {\bibinfo  {journal} {Nature Nanotechnology}\ }\textbf {\bibinfo
  {volume} {11}},\ \bibinfo {pages} {444} (\bibinfo {year} {2016})}\BibitemShut
  {NoStop}%
\bibitem [{\citenamefont {Boulle}\ \emph {et~al.}(2016)\citenamefont {Boulle},
  \citenamefont {Vogel}, \citenamefont {Yang}, \citenamefont {Pizzini},
  \citenamefont {de~Souza~Chaves}, \citenamefont {Locatelli}, \citenamefont
  {Mente{\c{s}}}, \citenamefont {Sala}, \citenamefont {Buda-Prejbeanu},
  \citenamefont {Klein} \emph {et~al.}}]{boulle2016room}%
  \BibitemOpen
  \bibfield  {author} {\bibinfo {author} {\bibfnamefont {O.}~\bibnamefont
  {Boulle}}, \bibinfo {author} {\bibfnamefont {J.}~\bibnamefont {Vogel}},
  \bibinfo {author} {\bibfnamefont {H.}~\bibnamefont {Yang}}, \bibinfo {author}
  {\bibfnamefont {S.}~\bibnamefont {Pizzini}}, \bibinfo {author} {\bibfnamefont
  {D.}~\bibnamefont {de~Souza~Chaves}}, \bibinfo {author} {\bibfnamefont
  {A.}~\bibnamefont {Locatelli}}, \bibinfo {author} {\bibfnamefont {T.~O.}\
  \bibnamefont {Mente{\c{s}}}}, \bibinfo {author} {\bibfnamefont
  {A.}~\bibnamefont {Sala}}, \bibinfo {author} {\bibfnamefont {L.~D.}\
  \bibnamefont {Buda-Prejbeanu}}, \bibinfo {author} {\bibfnamefont
  {O.}~\bibnamefont {Klein}},  \emph {et~al.},\ }\href@noop {} {\bibfield
  {journal} {\bibinfo  {journal} {Nature Nanotechnology}\ }\textbf {\bibinfo
  {volume} {11}},\ \bibinfo {pages} {449} (\bibinfo {year} {2016})}\BibitemShut
  {NoStop}%
\bibitem [{\citenamefont {Woo}\ \emph {et~al.}(2016)\citenamefont {Woo},
  \citenamefont {Litzius}, \citenamefont {Kr{\"u}ger}, \citenamefont {Im},
  \citenamefont {Caretta}, \citenamefont {Richter}, \citenamefont {Mann},
  \citenamefont {Krone}, \citenamefont {Reeve}, \citenamefont {Weigand} \emph
  {et~al.}}]{woo2016observation}%
  \BibitemOpen
  \bibfield  {author} {\bibinfo {author} {\bibfnamefont {S.}~\bibnamefont
  {Woo}}, \bibinfo {author} {\bibfnamefont {K.}~\bibnamefont {Litzius}},
  \bibinfo {author} {\bibfnamefont {B.}~\bibnamefont {Kr{\"u}ger}}, \bibinfo
  {author} {\bibfnamefont {M.-Y.}\ \bibnamefont {Im}}, \bibinfo {author}
  {\bibfnamefont {L.}~\bibnamefont {Caretta}}, \bibinfo {author} {\bibfnamefont
  {K.}~\bibnamefont {Richter}}, \bibinfo {author} {\bibfnamefont
  {M.}~\bibnamefont {Mann}}, \bibinfo {author} {\bibfnamefont {A.}~\bibnamefont
  {Krone}}, \bibinfo {author} {\bibfnamefont {R.~M.}\ \bibnamefont {Reeve}},
  \bibinfo {author} {\bibfnamefont {M.}~\bibnamefont {Weigand}},  \emph
  {et~al.},\ }\href@noop {} {\bibfield  {journal} {\bibinfo  {journal} {Nature
  Materials}\ }\textbf {\bibinfo {volume} {15}},\ \bibinfo {pages} {501–506}
  (\bibinfo {year} {2016})}\BibitemShut {NoStop}%
\bibitem [{\citenamefont {Sampaio}\ \emph {et~al.}(2013)\citenamefont
  {Sampaio}, \citenamefont {Cros}, \citenamefont {Rohart}, \citenamefont
  {Thiaville},\ and\ \citenamefont {Fert}}]{sampaio2013nucleation}%
  \BibitemOpen
  \bibfield  {author} {\bibinfo {author} {\bibfnamefont {J.}~\bibnamefont
  {Sampaio}}, \bibinfo {author} {\bibfnamefont {V.}~\bibnamefont {Cros}},
  \bibinfo {author} {\bibfnamefont {S.}~\bibnamefont {Rohart}}, \bibinfo
  {author} {\bibfnamefont {A.}~\bibnamefont {Thiaville}}, \ and\ \bibinfo
  {author} {\bibfnamefont {A.}~\bibnamefont {Fert}},\ }\href@noop {} {\bibfield
   {journal} {\bibinfo  {journal} {Nature Nanotechnology}\ }\textbf {\bibinfo
  {volume} {8}},\ \bibinfo {pages} {839} (\bibinfo {year} {2013})}\BibitemShut
  {NoStop}%
\bibitem [{\citenamefont {Ivanov}\ \emph {et~al.}(1999)\citenamefont {Ivanov},
  \citenamefont {Murav’ev},\ and\ \citenamefont {Sheka}}]{ivanov1999soliton}%
  \BibitemOpen
  \bibfield  {author} {\bibinfo {author} {\bibfnamefont {B.~A.}\ \bibnamefont
  {Ivanov}}, \bibinfo {author} {\bibfnamefont {V.}~\bibnamefont {Murav’ev}},
  \ and\ \bibinfo {author} {\bibfnamefont {D.~D.}\ \bibnamefont {Sheka}},\
  }\href@noop {} {\bibfield  {journal} {\bibinfo  {journal} {Journal of
  Experimental and Theoretical Physics}\ }\textbf {\bibinfo {volume} {89}},\
  \bibinfo {pages} {583} (\bibinfo {year} {1999})}\BibitemShut {NoStop}%
\bibitem [{\citenamefont {Sheka}\ \emph {et~al.}(2001)\citenamefont {Sheka},
  \citenamefont {Ivanov},\ and\ \citenamefont {Mertens}}]{sheka2001internal}%
  \BibitemOpen
  \bibfield  {author} {\bibinfo {author} {\bibfnamefont {D.~D.}\ \bibnamefont
  {Sheka}}, \bibinfo {author} {\bibfnamefont {B.~A.}\ \bibnamefont {Ivanov}}, \
  and\ \bibinfo {author} {\bibfnamefont {F.~G.}\ \bibnamefont {Mertens}},\
  }\href@noop {} {\bibfield  {journal} {\bibinfo  {journal} {Physical Review
  B}\ }\textbf {\bibinfo {volume} {64}},\ \bibinfo {pages} {024432} (\bibinfo
  {year} {2001})}\BibitemShut {NoStop}%
\bibitem [{\citenamefont {Kim}\ \emph {et~al.}(2014)\citenamefont {Kim},
  \citenamefont {Garcia-Sanchez}, \citenamefont {Sampaio}, \citenamefont
  {Moreau-Luchaire}, \citenamefont {Cros},\ and\ \citenamefont {Fert}}]{44}%
  \BibitemOpen
  \bibfield  {author} {\bibinfo {author} {\bibfnamefont {J.~V.}\ \bibnamefont
  {Kim}}, \bibinfo {author} {\bibfnamefont {F.}~\bibnamefont {Garcia-Sanchez}},
  \bibinfo {author} {\bibfnamefont {J.}~\bibnamefont {Sampaio}}, \bibinfo
  {author} {\bibfnamefont {C.}~\bibnamefont {Moreau-Luchaire}}, \bibinfo
  {author} {\bibfnamefont {V.}~\bibnamefont {Cros}}, \ and\ \bibinfo {author}
  {\bibfnamefont {A.}~\bibnamefont {Fert}},\ }\href {\doibase
  10.1103/PhysRevB.90.064410} {\bibfield  {journal} {\bibinfo  {journal} {Phys.
  Rev. B}\ }\textbf {\bibinfo {volume} {90}},\ \bibinfo {pages} {064410}
  (\bibinfo {year} {2014})}\BibitemShut {NoStop}%
\bibitem [{\citenamefont {Iwasaki}\ \emph {et~al.}(2013)\citenamefont
  {Iwasaki}, \citenamefont {Mochizuki},\ and\ \citenamefont {Nagaosa}}]{69}%
  \BibitemOpen
  \bibfield  {author} {\bibinfo {author} {\bibfnamefont {J.}~\bibnamefont
  {Iwasaki}}, \bibinfo {author} {\bibfnamefont {M.}~\bibnamefont {Mochizuki}},
  \ and\ \bibinfo {author} {\bibfnamefont {N.}~\bibnamefont {Nagaosa}},\
  }\href@noop {} {\bibfield  {journal} {\bibinfo  {journal} {Nature
  Nanotechnology}\ }\textbf {\bibinfo {volume} {8}},\ \bibinfo {pages}
  {742–747} (\bibinfo {year} {2013})}\BibitemShut {NoStop}%
\bibitem [{\citenamefont {Gareeva}\ and\ \citenamefont
  {Guslienko}(2016)}]{gareeva2016magnetic}%
  \BibitemOpen
  \bibfield  {author} {\bibinfo {author} {\bibfnamefont {Z.~V.}\ \bibnamefont
  {Gareeva}}\ and\ \bibinfo {author} {\bibfnamefont {K.~Y.}\ \bibnamefont
  {Guslienko}},\ }\href@noop {} {\bibfield  {journal} {\bibinfo  {journal}
  {Physica Status Solidi (RRL)-Rapid Research Letters}\ }\textbf {\bibinfo
  {volume} {10}},\ \bibinfo {pages} {227} (\bibinfo {year} {2016})}\BibitemShut
  {NoStop}%
\bibitem [{\citenamefont {Guslienko}\ and\ \citenamefont
  {Gareeva}(2016)}]{guslienko2016gyrotropic}%
  \BibitemOpen
  \bibfield  {author} {\bibinfo {author} {\bibfnamefont {K.~Y.}\ \bibnamefont
  {Guslienko}}\ and\ \bibinfo {author} {\bibfnamefont {Z.~V.}\ \bibnamefont
  {Gareeva}},\ }\href@noop {} {\bibfield  {journal} {\bibinfo  {journal} {IEEE
  Magn. Lett., accepted}\ } (\bibinfo {year} {2016})}\BibitemShut {NoStop}%
\bibitem [{\citenamefont {Mruczkiewicz}\ \emph {et~al.}(2016)\citenamefont
  {Mruczkiewicz}, \citenamefont {Gruszecki}, \citenamefont {Zelent},\ and\
  \citenamefont {Krawczyk}}]{mruczkiewicz2016collective}%
  \BibitemOpen
  \bibfield  {author} {\bibinfo {author} {\bibfnamefont {M.}~\bibnamefont
  {Mruczkiewicz}}, \bibinfo {author} {\bibfnamefont {P.}~\bibnamefont
  {Gruszecki}}, \bibinfo {author} {\bibfnamefont {M.}~\bibnamefont {Zelent}}, \
  and\ \bibinfo {author} {\bibfnamefont {M.}~\bibnamefont {Krawczyk}},\
  }\href@noop {} {\bibfield  {journal} {\bibinfo  {journal} {Physical Review
  B}\ }\textbf {\bibinfo {volume} {93}},\ \bibinfo {pages} {174429} (\bibinfo
  {year} {2016})}\BibitemShut {NoStop}%
\bibitem [{\citenamefont {Makhfudz}\ \emph {et~al.}(2012)\citenamefont
  {Makhfudz}, \citenamefont {Kr{\"u}ger},\ and\ \citenamefont
  {Tchernyshyov}}]{makhfudz2012inertia}%
  \BibitemOpen
  \bibfield  {author} {\bibinfo {author} {\bibfnamefont {I.}~\bibnamefont
  {Makhfudz}}, \bibinfo {author} {\bibfnamefont {B.}~\bibnamefont
  {Kr{\"u}ger}}, \ and\ \bibinfo {author} {\bibfnamefont {O.}~\bibnamefont
  {Tchernyshyov}},\ }\href@noop {} {\bibfield  {journal} {\bibinfo  {journal}
  {Physical Review Letters}\ }\textbf {\bibinfo {volume} {109}},\ \bibinfo
  {pages} {217201} (\bibinfo {year} {2012})}\BibitemShut {NoStop}%
\bibitem [{\citenamefont {Castel}\ \emph {et~al.}(2012)\citenamefont {Castel},
  \citenamefont {Youssef}, \citenamefont {Boust}, \citenamefont {Weil},
  \citenamefont {Pigeau}, \citenamefont {de~Loubens}, \citenamefont {Naletov},
  \citenamefont {Klein},\ and\ \citenamefont
  {Vukadinovic}}]{castel2012perpendicular}%
  \BibitemOpen
  \bibfield  {author} {\bibinfo {author} {\bibfnamefont {V.}~\bibnamefont
  {Castel}}, \bibinfo {author} {\bibfnamefont {J.~B.}\ \bibnamefont {Youssef}},
  \bibinfo {author} {\bibfnamefont {F.}~\bibnamefont {Boust}}, \bibinfo
  {author} {\bibfnamefont {R.}~\bibnamefont {Weil}}, \bibinfo {author}
  {\bibfnamefont {B.}~\bibnamefont {Pigeau}}, \bibinfo {author} {\bibfnamefont
  {G.}~\bibnamefont {de~Loubens}}, \bibinfo {author} {\bibfnamefont
  {V.}~\bibnamefont {Naletov}}, \bibinfo {author} {\bibfnamefont
  {O.}~\bibnamefont {Klein}}, \ and\ \bibinfo {author} {\bibfnamefont
  {N.}~\bibnamefont {Vukadinovic}},\ }\href@noop {} {\bibfield  {journal}
  {\bibinfo  {journal} {Physical Review B}\ }\textbf {\bibinfo {volume} {85}},\
  \bibinfo {pages} {184419} (\bibinfo {year} {2012})}\BibitemShut {NoStop}%
\bibitem [{\citenamefont {Taurel}\ \emph {et~al.}(2016)\citenamefont {Taurel},
  \citenamefont {Valet}, \citenamefont {Naletov}, \citenamefont {Vukadinovic},
  \citenamefont {de~Loubens},\ and\ \citenamefont
  {Klein}}]{taurel2016complete}%
  \BibitemOpen
  \bibfield  {author} {\bibinfo {author} {\bibfnamefont {B.}~\bibnamefont
  {Taurel}}, \bibinfo {author} {\bibfnamefont {T.}~\bibnamefont {Valet}},
  \bibinfo {author} {\bibfnamefont {V.~V.}\ \bibnamefont {Naletov}}, \bibinfo
  {author} {\bibfnamefont {N.}~\bibnamefont {Vukadinovic}}, \bibinfo {author}
  {\bibfnamefont {G.}~\bibnamefont {de~Loubens}}, \ and\ \bibinfo {author}
  {\bibfnamefont {O.}~\bibnamefont {Klein}},\ }\href@noop {} {\bibfield
  {journal} {\bibinfo  {journal} {Physical Review B}\ }\textbf {\bibinfo
  {volume} {93}},\ \bibinfo {pages} {184427} (\bibinfo {year}
  {2016})}\BibitemShut {NoStop}%
\bibitem [{\citenamefont {Vansteenkiste}\ \emph {et~al.}(2014)\citenamefont
  {Vansteenkiste}, \citenamefont {Leliaert}, \citenamefont {Dvornik},
  \citenamefont {Helsen}, \citenamefont {Garcia-Sanchez},\ and\ \citenamefont
  {Van~Waeyenberge}}]{4899186}%
  \BibitemOpen
  \bibfield  {author} {\bibinfo {author} {\bibfnamefont {A.}~\bibnamefont
  {Vansteenkiste}}, \bibinfo {author} {\bibfnamefont {J.}~\bibnamefont
  {Leliaert}}, \bibinfo {author} {\bibfnamefont {M.}~\bibnamefont {Dvornik}},
  \bibinfo {author} {\bibfnamefont {M.}~\bibnamefont {Helsen}}, \bibinfo
  {author} {\bibfnamefont {F.}~\bibnamefont {Garcia-Sanchez}}, \ and\ \bibinfo
  {author} {\bibfnamefont {B.}~\bibnamefont {Van~Waeyenberge}},\ }\href
  {\doibase http://dx.doi.org/10.1063/1.4899186} {\bibfield  {journal}
  {\bibinfo  {journal} {AIP Advances}\ }\textbf {\bibinfo {volume} {4}},\
  \bibinfo {eid} {107133} (\bibinfo {year} {2014})}\BibitemShut {NoStop}%
\bibitem [{\citenamefont {Bogdanov}\ and\ \citenamefont
  {R{\"o}{\ss}ler}(2001)}]{bogdanov2001chiral}%
  \BibitemOpen
  \bibfield  {author} {\bibinfo {author} {\bibfnamefont {A.}~\bibnamefont
  {Bogdanov}}\ and\ \bibinfo {author} {\bibfnamefont {U.}~\bibnamefont
  {R{\"o}{\ss}ler}},\ }\href@noop {} {\bibfield  {journal} {\bibinfo  {journal}
  {Physical Review Letters}\ }\textbf {\bibinfo {volume} {87}},\ \bibinfo
  {pages} {037203} (\bibinfo {year} {2001})}\BibitemShut {NoStop}%
\bibitem [{\citenamefont {Nakatani}\ \emph {et~al.}(2016)\citenamefont
  {Nakatani}, \citenamefont {Hayashi}, \citenamefont {Kanai}, \citenamefont
  {Fukami},\ and\ \citenamefont {Ohno}}]{nakatani2016electric}%
  \BibitemOpen
  \bibfield  {author} {\bibinfo {author} {\bibfnamefont {Y.}~\bibnamefont
  {Nakatani}}, \bibinfo {author} {\bibfnamefont {M.}~\bibnamefont {Hayashi}},
  \bibinfo {author} {\bibfnamefont {S.}~\bibnamefont {Kanai}}, \bibinfo
  {author} {\bibfnamefont {S.}~\bibnamefont {Fukami}}, \ and\ \bibinfo {author}
  {\bibfnamefont {H.}~\bibnamefont {Ohno}},\ }\href@noop {} {\bibfield
  {journal} {\bibinfo  {journal} {Applied Physics Letters}\ }\textbf {\bibinfo
  {volume} {108}},\ \bibinfo {pages} {152403} (\bibinfo {year}
  {2016})}\BibitemShut {NoStop}%
\bibitem [{\citenamefont {Natarajarathinam}\ \emph {et~al.}(2012)\citenamefont
  {Natarajarathinam}, \citenamefont {Tadisina}, \citenamefont {Mewes},
  \citenamefont {Watts}, \citenamefont {Chen},\ and\ \citenamefont
  {Gupta}}]{Natarajarathinam12}%
  \BibitemOpen
  \bibfield  {author} {\bibinfo {author} {\bibfnamefont {A.}~\bibnamefont
  {Natarajarathinam}}, \bibinfo {author} {\bibfnamefont {Z.~R.}\ \bibnamefont
  {Tadisina}}, \bibinfo {author} {\bibfnamefont {T.}~\bibnamefont {Mewes}},
  \bibinfo {author} {\bibfnamefont {S.}~\bibnamefont {Watts}}, \bibinfo
  {author} {\bibfnamefont {E.}~\bibnamefont {Chen}}, \ and\ \bibinfo {author}
  {\bibfnamefont {S.}~\bibnamefont {Gupta}},\ }\href@noop {} {\bibfield
  {journal} {\bibinfo  {journal} {Journal of Applied Physics}\ }\textbf
  {\bibinfo {volume} {112}},\ \bibinfo {pages} {053909} (\bibinfo {year}
  {2012})}\BibitemShut {NoStop}%
\bibitem [{\citenamefont {Yu}\ \emph {et~al.}(2012)\citenamefont {Yu},
  \citenamefont {Huber}, \citenamefont {Schwarze}, \citenamefont {Brandl},
  \citenamefont {Rapp}, \citenamefont {Berberich}, \citenamefont {Duerr},\ and\
  \citenamefont {Grundler}}]{Yu12}%
  \BibitemOpen
  \bibfield  {author} {\bibinfo {author} {\bibfnamefont {H.}~\bibnamefont
  {Yu}}, \bibinfo {author} {\bibfnamefont {R.}~\bibnamefont {Huber}}, \bibinfo
  {author} {\bibfnamefont {T.}~\bibnamefont {Schwarze}}, \bibinfo {author}
  {\bibfnamefont {F.}~\bibnamefont {Brandl}}, \bibinfo {author} {\bibfnamefont
  {T.}~\bibnamefont {Rapp}}, \bibinfo {author} {\bibfnamefont {P.}~\bibnamefont
  {Berberich}}, \bibinfo {author} {\bibfnamefont {G.}~\bibnamefont {Duerr}}, \
  and\ \bibinfo {author} {\bibfnamefont {D.}~\bibnamefont {Grundler}},\
  }\href@noop {} {\bibfield  {journal} {\bibinfo  {journal} {Applied Physics
  Letters}\ }\textbf {\bibinfo {volume} {100}},\ \bibinfo {pages} {262412}
  (\bibinfo {year} {2012})}\BibitemShut {NoStop}%
\bibitem [{\citenamefont {Nagaosa}\ and\ \citenamefont
  {Tokura}(2013)}]{nagaosa2013topological}%
  \BibitemOpen
  \bibfield  {author} {\bibinfo {author} {\bibfnamefont {N.}~\bibnamefont
  {Nagaosa}}\ and\ \bibinfo {author} {\bibfnamefont {Y.}~\bibnamefont
  {Tokura}},\ }\href@noop {} {\bibfield  {journal} {\bibinfo  {journal} {Nature
  Nanotechnology}\ }\textbf {\bibinfo {volume} {8}},\ \bibinfo {pages} {899}
  (\bibinfo {year} {2013})}\BibitemShut {NoStop}%
\bibitem [{\citenamefont {Guslienko}(2015)}]{7061384}%
  \BibitemOpen
  \bibfield  {author} {\bibinfo {author} {\bibfnamefont {K.}~\bibnamefont
  {Guslienko}},\ }\href {\doibase 10.1109/LMAG.2015.2413758} {\bibfield
  {journal} {\bibinfo  {journal} {Magnetics Letters, IEEE}\ }\textbf {\bibinfo
  {volume} {6}},\ \bibinfo {pages} {4000104} (\bibinfo {year}
  {2015})}\BibitemShut {NoStop}%
\bibitem [{\citenamefont {Kisielewski}\ \emph {et~al.}(2003)\citenamefont
  {Kisielewski}, \citenamefont {Maziewski}, \citenamefont {Zablotskii},
  \citenamefont {Polyakova}, \citenamefont {Garcia}, \citenamefont {Wawro},\
  and\ \citenamefont {Baczewski}}]{kisielewski2003drastic}%
  \BibitemOpen
  \bibfield  {author} {\bibinfo {author} {\bibfnamefont {M.}~\bibnamefont
  {Kisielewski}}, \bibinfo {author} {\bibfnamefont {A.}~\bibnamefont
  {Maziewski}}, \bibinfo {author} {\bibfnamefont {V.}~\bibnamefont
  {Zablotskii}}, \bibinfo {author} {\bibfnamefont {T.}~\bibnamefont
  {Polyakova}}, \bibinfo {author} {\bibfnamefont {J.}~\bibnamefont {Garcia}},
  \bibinfo {author} {\bibfnamefont {A.}~\bibnamefont {Wawro}}, \ and\ \bibinfo
  {author} {\bibfnamefont {L.}~\bibnamefont {Baczewski}},\ }\href@noop {}
  {\bibfield  {journal} {\bibinfo  {journal} {Journal of Applied Physics}\
  }\textbf {\bibinfo {volume} {93}},\ \bibinfo {pages} {6966} (\bibinfo {year}
  {2003})}\BibitemShut {NoStop}%
\bibitem [{\citenamefont {Virot}\ \emph {et~al.}(2012)\citenamefont {Virot},
  \citenamefont {Favre}, \citenamefont {Hayn},\ and\ \citenamefont
  {Kuz'min}}]{virot2012theory}%
  \BibitemOpen
  \bibfield  {author} {\bibinfo {author} {\bibfnamefont {F.}~\bibnamefont
  {Virot}}, \bibinfo {author} {\bibfnamefont {L.}~\bibnamefont {Favre}},
  \bibinfo {author} {\bibfnamefont {R.}~\bibnamefont {Hayn}}, \ and\ \bibinfo
  {author} {\bibfnamefont {M.}~\bibnamefont {Kuz'min}},\ }\href@noop {}
  {\bibfield  {journal} {\bibinfo  {journal} {Journal of Physics D: Applied
  Physics}\ }\textbf {\bibinfo {volume} {45}},\ \bibinfo {pages} {405003}
  (\bibinfo {year} {2012})}\BibitemShut {NoStop}%
\bibitem [{\citenamefont {Mochizuki}(2012{\natexlab{b}})}]{mochizuki2012spin}%
  \BibitemOpen
  \bibfield  {author} {\bibinfo {author} {\bibfnamefont {M.}~\bibnamefont
  {Mochizuki}},\ }\href@noop {} {\bibfield  {journal} {\bibinfo  {journal}
  {Physical Review Letters}\ }\textbf {\bibinfo {volume} {108}},\ \bibinfo
  {pages} {017601} (\bibinfo {year} {2012}{\natexlab{b}})}\BibitemShut
  {NoStop}%
\end{thebibliography}%

\end{document}